# Time evolution of photon-pulse propagation in scattering and absorbing media: the Dynamic Radiative Transfer System.

A. Georgakopoulos*,[a], K. Politopoulos [a] and E. Georgiou[b].

[a]Biomedical Optics & Applied Biophysics Laboratory, School of Electrical and Computer Engineering, National Technical University of Athens, Heroon Polytechniou 9, 15780 Athens, Greece
[b]Department of Electrical Engineering, Technological Educational Institute of Crete, 71500 Stavromenos, Heraklion, Greece

*Corresponding Author: Tel: +30-2107723048, Fax: +30-2107723894
E-mail address: tassosgeo@central.ntua.gr (A. Georgakopoulos)

## Abstract

A new dynamic-system approach to the problem of radiative transfer inside scattering and absorbing media is presented, directly based on first-hand physical principles. This method, the Dynamic Radiative Transfer System (DRTS), employs a dynamical system formality using a global sparse matrix, which characterizes the physical, optical and geometrical properties of the material-volume of interest. The new system state is generated by the above time-independent matrix, using simple matrix-vector multiplication for each subsequent time step. DRTS is capable of calculating accurately the time evolution of photon propagation in media of complex structure and shape. The flexibility of DRTS allows the integration of time-dependent sources, boundary conditions, different media and several optical phenomena like reflection and refraction in a unified and consistent way. Various examples of DRTS simulation results are presented for ultra-fast light pulse 3-D propagation, demonstrating greatly reduced computational cost and resource requirements compared to other methods.

## 1. Introduction

Radiative transfer models attempt to describe the temporal and spatial behavior of photon propagation through scattering and absorbing media, which is important in many scientific areas such as astrophysics [1–3], oceanography, high energy physics (neutrino transport)[4], thermal transfer [5–7][39], image processing and, in recent years, bio-sciences [8–11]. Starting from a conceptual framework for a simple description of the physical phenomenon, the well-known "Radiative Transfer Equation" (RTE) was developed:

$$\frac{1}{c}\frac{\partial I(\boldsymbol{x},\boldsymbol{s},t)}{\partial t} + \frac{\partial I(\boldsymbol{x},\boldsymbol{s},t)}{\partial s} + (\mu_a + \mu_s)I(\boldsymbol{x},\boldsymbol{s},t) - \mu_s \iint I(\boldsymbol{x},\boldsymbol{s}',t)g(\boldsymbol{s},\boldsymbol{s}')d\boldsymbol{s}' = 0 \quad (1)$$

Many different solutions or approximations for RTE have been proposed [12], such as Monte Carlo [13,14], Diffusion Approximation, calculations using different basis functions [4,6,15–17], Finite-Element-like methods (FEM) [1,5,7] and combinations of the above [5,8,10,18,19]. In a very simple case a solution has been found based on a "moving'' unity partition of normal distribution, decreasing over time by an exponential factor according to the absorption parameter. Diffusion approximation connects the directed flow with undirected flow by a constant that simplifies the mathematical problem and turns it to an elliptical partial differential equation. However, this is not physically true near sources or boundaries. Another approach, use spherical harmonics [4,6,20,21], does not lead to an applicable solution because of the symmetry of base functions that makes almost impossible to represent a directional flow. Mathematical approximations based on finite element methods (FEM) have been proposed by many authors [1,5,7,22], treating complicated structures such as the human body with partitioning of Euclidean space and creation of mesh structures. These methods however do not preserve the positivity of intensity and produce potential instabilities because of the mathematical type of the RTE model. A combination of FEM with Diffusion approximation reduces the instability but does not fully resolve the problem [8,18,19].

Because of mathematical difficulties [20,21,23,24], procedures based on Mode Carlo (MC) methods are believed to be the best practice, since calculations are going down to physical processes. Parallel array processors have been used [25,26] to handle the huge amount of calculations required by this type of methods. Usually in standard MC techniques only space partition is implemented, so there is a lack of time and direction information and these are not suitable to describe transient phenomena. Specifically, since only spatial information is given, there is no practical way to convolve the results with any specific pulsed-shaped source. In time-dependent MC [27,28] the computational requirements explode, especially in systems with dense partitioning in space and time, as this method generally requires computational effort proportional to the dynamic range of the expected values – here on the order of $10^{20}$. These MC shortcomings are particularly evident in cases involving ultra-short light pulses.

A solution of the time-dependent photon transport has been attempted by Fourier transform [8,23], "Time-dependent Monte-Carlo"[29,30] or other time-domain methods [31–33,40-42]. These can lead to a meaningful approximate solution only when source power changes are slow enough compared to the fast rates of scattering effects. Under these circumstances, intensity will represent the mean-time value under the effect of power-source changes. But when source changes are fast, frequencies of the source will be convolved with those of the system, a fact that makes computations impractical for very short time-scale phenomena, due to the range of frequencies involved, which tend to infinity. Recent developments in medical diagnostics, however, using $fs$ laser pulses, e.g. for internal tumor imaging, will require accurate solutions of the radiation propagation problem in absorptive-diffusive media in very short time scales.

There are many difficulties regarding numerical solution of the RTE, or its physical incompleteness or inconsistencies [24,34,35]. To overcome these problems, a new method is described in this paper, based on the physical photon-matter interaction in a consistent treatment. This method, the Dynamic Radiative Transfer

System (DRTS), computes the time evolution of photon distribution of inside scattering absorbing media by the construction of a probabilistic dynamic system. This is especially useful in bio-engineering for e.g. solving the Optical Tomography "forward problem" [36] using pulsed excitation. . In such cases DRTS is quite capable to deliver accurate results, where other methods might be questionable. DRTS is also capable of calculating accurately the time evolution of photon propagation in media of complex structure and shape. Additionally, DRTS allows the integration of time-dependent sources, boundary conditions, different media and several optical phenomena (like reflection and refraction) in a unified and consistent way. Especially for the purpose of instrument design, the new DRTS method provides a tool that allows an accurate measurement-system implementation. Therefore DRTS can be viewed as an improved and generalized solution method, accurately treating time effects in both short and long time scale, which avoids the inconsistencies and drawbacks of the RTE model.

In the next section the physical phenomenon will be expressed as a dynamical system and the computation of the system matrix elements will be detailed. Time evolution of the photon propagation can afterwards be computed by the simplest linear algebra operations (matrix-vector multiplication and vector additions). Then, the algorithm based on the theoretical analysis is constructed and simulation results are initially demonstrated for several cases of isotropic media with different optical properties. In the last section, a thorough comparative analysis of DRTS vs. the other methods will be detailed.

## 2. Theoretical foundation of photon propagation

The Dynamic Radiative Transfer System (DRTS) is a new method that models photon propagation through scattering and absorbing media, applicable to time intervals down to the order of femtosecond.

Absorption and scattering processes are considered independent. There are three measures that characterize absorption and scattering probabilities, the absorption parameter $\mu_a$ the scattering parameter $\mu_s$ and the probability kernel $g(\mathbf{S}, \mathbf{S}')$ expressing the directional change. The first two are measures over the path of photons that travel inside the medium. The third one expresses the conditional possibility that a photon changes direction from $\mathbf{S}$ to $\mathbf{S}'$ if it has been scattered.

By definition $1/\mu_s$ expresses the mean value of the exponential probability for the photons to be subjected to scattering inside the medium. The expected population of photons that are scattered ($N^S$) during a time step $\Delta t$ can be derived by the cumulative distribution function (CDF) of the exponential distribution:

$$N^S(t + \Delta t) = N(t) P(\text{photon scattering during time interval } \Delta t) \Rightarrow$$

$$N^S(t + \Delta t) = N(t)\left(1 - e^{-\mu_s c \Delta t}\right) \tag{2}$$

Correspondingly the photon population that is not subjected to any scattering event ($N^0$), during the time interval $\Delta t$, will be:

$$N^{S^0}(t + \Delta t) = N(t) - N^S(t + \Delta t) = N(t)e^{-\mu_s c \Delta t} \tag{3}$$

The probability that a photon is subjected to $k$ scattering events in $\Delta t$ can be derived by the Poisson distribution with expected value $\mu_s c \Delta t$. The population of scattered photons can be grouped in different populations $N^{S^0}, N^{S^1}, N^{S^2}, ...$ according to the number of the scattering events encountered, so:

$$N^{S^i}(t + \Delta t) = N(t)P(i = scatter\ events) = N(t)\frac{e^{-\mu_s c \Delta t}(\mu_s c \Delta t)^i}{i!} \tag{4}$$

From the above equation, it can be proved that for very small time intervals the probability of the photon population to be subjected to more than one scattering event is negligibly small. This can be proved by calculating the following limit, for $i > 1$:

$$\lim_{\Delta t \to 0} \frac{N^{S^i}}{N^{S^1}} = \lim_{\Delta t \to 0} \left( \frac{\frac{e^{-\mu_s c \Delta t}(\mu_s c \Delta t)^i}{i!}}{e^{-\mu_s c \Delta t}} \right) = \lim_{\Delta t \to 0} \frac{(\mu_s c \Delta t)^i}{i!} = 0, \quad \text{when} \quad \mu_s c \Delta t \ll 1 \quad (5)$$

At this point it might appear that only single-scattering events are considered. According to Eq. (5) as $\Delta t$ gets smaller, the probability of multiple scattering is exponentially reduced. However, our method has the capability of including higher-order scattering events during this interval $\Delta t$, if so desired, as it will be detailed later in section 3, after the construction of the system matrix A. Essentially multiple scattering events are produced as a combinatorial sequence of single-scatterings.

To conserve probability and therefore energy, while ignoring multiple scattering events, we approximate the probability of one scattering event $P(i = 1)$ as $1 - P(i = 0)$. Thus, regarding the scattering process we can split the photon population in two groups $N^0$ and $N^1$:

$$N^{S^0}(t + \Delta t) = N(t)e^{-\mu_s c \Delta t} \quad (6)$$

$$N^{S^1}(t + \Delta t) = N(t)(1 - e^{-\mu_s c \Delta t}), \quad \text{when} \quad \mu_s c \Delta t \ll 1 \quad (7)$$

For the absorption process, by definition $1/\mu_a$ expresses the mean value of the exponential probability for the decay of photons that travel inside the medium. The expected population ($N$) of photons surviving after a small time step $\Delta t$ can be derived by the CDF of the exponential distribution:

$$N(t + \Delta t) = N(t)(1 - P(absorption\ during\ time\ interval\ \Delta t)) = N(t)e^{-\mu_a c \Delta t} \quad (8)$$

By combining both scattering and absorption processes, the expected population group of photons surviving after a small time step $\Delta t$ will be:

$$N^0(t + \Delta t) = N(t)e^{-\mu_s c \Delta t}e^{-\mu_a c \Delta t} \quad (9)$$

$$N^1(t + \Delta t) = N(t)(1 - e^{-\mu_s c \Delta t})e^{-\mu_a c \Delta t}, \quad \text{when} \quad \mu_s c \Delta t \ll 1 \quad (10)$$

Notice that from Eq. (9) and henceforth for the sake of notation simplicity, we use the superscripts 0,1,…, to denote in all cases the surviving population after 0,1,…, scattering events.

The previous equations give information about the partition of photon population according to the events that occur, not including yet any positional-directional information. Below we calculate the position and direction of photons at the new time-point $t + \Delta t$.

Let $N(\mathbf{X}, \mathbf{S}, t)$ be the photon population in a spherical volume $dV$, radius $dr$ centered at point $\mathbf{X}$, travelling at direction $\mathbf{S}$ with speed $c$ at time $t$.

The part of the population $N(\mathbf{X}, \mathbf{S}, t)$ subject to no scattering event (type $N^0$ from Eq. (9)) will move after time $\Delta t$ to an equal sphere at a new position $\mathbf{B}$:

$$N^0(\mathbf{B}, \mathbf{S}, t + \Delta t) = e^{-\mu_a c \Delta t}e^{-\mu_s c \Delta t}N(\mathbf{X}, \mathbf{S}, t)), \quad \text{where} \quad \mathbf{B} = \mathbf{X} + c\Delta t \mathbf{S} \quad (11)$$

For the scattered population, in order to calculate the new position – direction, we start by considering the possible trajectories of photons starting from a single point towards an initial direction.

Let some photons at time $t$ be at point $X$ and travel at direction $S$, which during time $\Delta t$ are scattered into direction $S'$. These photons will end-up at time $t + \Delta t$ at any point $X'$ along the line segment $AB$ under the condition that only one change of direction from $S$ to $S'$ has been done, where $A = X + c\Delta t S'$ and $B = X + c\Delta t S$ as shown in *Figure 1*. Point $A$ is defined by the position of the photons that changed direction at time t, while point $B$ is defined by the position of photons that changed direction at time $t + \Delta t$.

All possible trajectories of photons will be either on direction $S$ or $S'$ so all events will be on the plane $S \times S'$, also all photons cannot exit the sphere $(X, c\Delta t)$ because of constant light speed. Now let the scattering event occur at a moment between $(t, t + \Delta t)$ at a point $C$ lying along the line segment $XB$. At time $t + \Delta t$ this photon will reach point $X'$. The triangle $CBX'$ is isosceles because of the following:

$$\|XC\| + \|CB\| = \|XC\| + \|CX'\| = c\Delta t \Rightarrow \|CB\| = \|CX'\| \tag{12}$$

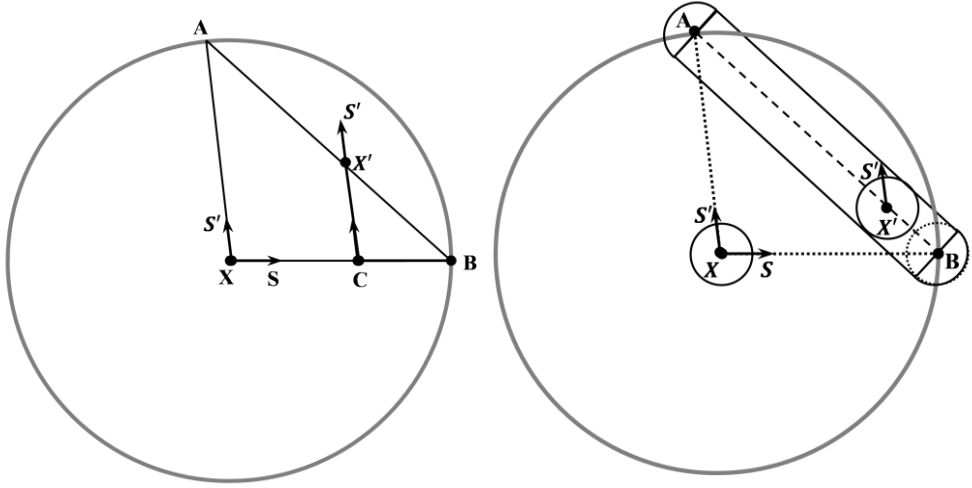

*Figure 1.* Photon trajectories starting from point $X$ after time $\Delta t$ are contained within the large sphere of radius $c\Delta t$. Photon population of type $N^0$ starting from sphere $(X, dr)$ will arrive at sphere $(B, dr)$. Population of type $N^1$ will change direction from $S$ to $S'$ at $C$ and arrive at time $t + \Delta t$ to a point $X'$ that lies on the line segment $AB$.

Since $S, S'$ are fixed so does the angle $X'CB$, so does the angle $CBX'$ which is equal to $XBA$. So the locus of the arrival points $X'$ will be the line segment $AB$. A function can be constructed that connects the directions $S$ and $S'$ for fixed time step $\Delta t$ and points $X, X'$.

This function can be computed solving the following system of vector equations.

$$\left.\begin{array}{r} A = X + c\Delta t S' \\ B = X + c\Delta t S \\ B = k(X' - A) + A \quad k \geq 0 \\ S, S' \text{ unit vectors} \end{array}\right\} \Rightarrow \left.\begin{array}{r} B - A = c\Delta t(S - S') \\ B - A = k(X' - X - c\Delta t S') \\ B - X = k(X' - X - c\Delta t S') + X + c\Delta t S' - X \end{array}\right\} \Rightarrow$$

$$\left.\begin{array}{r} c\Delta t(S - S') = k(X' - X - c\Delta t S') \\ \|B - X\|^2 = k^2\|(X' - X - c\Delta t S') + c\Delta t S'\|^2 = (c\Delta t)^2 \end{array}\right\} \Rightarrow$$

$$\left.\begin{array}{r} c\Delta t(S - S') = k(X' - X - c\Delta t S') \\ (c\Delta t)^2 = k^2\|(X' - X - c\Delta t S')\|^2 + 2k(X' - X - c\Delta t S') * (c\Delta t S') + (c\Delta t)^2\|S'\|^2 \end{array}\right\} \Rightarrow$$

$$\left.\begin{array}{r} c\Delta t(S - S') = k(X' - X - c\Delta t S') \\ k = \dfrac{-2(X' - X - c\Delta t S') * (c\Delta t S')}{\|(X' - X - c\Delta t S')\|^2} \end{array}\right\} \Rightarrow$$

$$S = S' - \frac{2(X' - X - c\Delta t S') * (X' - X - c\Delta t S') * S'}{\|(X' - X - c\Delta t S')\|^2} \tag{13}$$

Note that because of the symmetry of the equations, the inverse relation is easily calculated. So the function that connects the directions $S$ and $S'$ is one to one, given $X, X'$.

The above analysis reveals a very important general corollary regarding the seemingly random nature of the scattering process. Instead, the geometrical and physical (light-speed c) constraints demonstrate a subtle "interconnection" between allowed space points and/or directions. Specifically, starting from specific initial photon state $(X, S)$, within the circle $c\Delta t$, other points $X'$ can be reached by a single scattering event *only* via a specific direction change from $S$ to $S'$. Correspondingly, two directions can only be "connected" only via specific space points.

The previous geometrical analysis has established the possible arrival points of scattered photons emanating from a single point, after time $\Delta t$. In the next step the spatial distribution of population $N^1$ is calculated.

Change of direction has equal probability to occur over time $\Delta t$. So the population of photons with the new direction $S'$ will be uniformly distributed over the space around the line segment $AB$. The expected part of the population of photons that will change direction from $S$ to $S'$ is given by the kernel $g(S, S')$.

So the population $g(S, S')(1 - e^{-\mu_s c \Delta t})e^{-\mu_a c \Delta t}N(X, S, t)$ will be uniformly distributed around the segment $AB$. Let's define $p_{X,S}$, to be the density of photons in the volume around the line segment. Then we have:

$$\iiint_{V(AB,dr)} p_{X,S} dV = g(S, S')(1 - e^{-\mu_s c \Delta t})e^{-\mu_a c \Delta t}N(X, S, t) \Rightarrow \quad (14)$$

$$p_{X,S} \iiint_{V(AB,dr)} dV = g(S, S')(1 - e^{-\mu_s c \Delta t})e^{-\mu_a c \Delta t}N(X, S, t) \quad (15)$$

$p_{X,S}(cylinder\ Vol\ (AB, dr) + half\ spheres\ dr\ at\ points\ A, B) = (1 - e^{-\mu_s c \Delta t})e^{-\mu_a c \Delta t}N(X, S, t) \Rightarrow$

$$\left. \begin{aligned} p_{X,S}\left(\pi dr^2 \|AB\| + \frac{4}{3}\pi dr^3\right) &= N(X, S, t)g(S, S')(1 - e^{-\mu_s c \Delta t})e^{-\mu_a c \Delta t} \\ \|AB\| &= \sqrt{2}c\Delta t(1 - S * S') \end{aligned} \right\} \Rightarrow$$

$$p_{X,S} = \frac{g(S, S')(1 - e^{-\mu_s c \Delta t})e^{-\mu_a c \Delta t}}{\left(\pi dr^2 \sqrt{2}c\Delta t(1 - S * S') + \frac{4}{3}\pi dr^3\right)} N(X, S, t) \quad (16)$$

So, the number of photons inside a sphere $(X', dr)$ directed to $S'$ at time $t + \Delta t$, which originated from sphere $(X, dr)$ directed to $S$ will be:

$$N^1\big((X', S', t + \Delta t) \leftarrow (X, S, t)\big) = p_{X,S} \cdot dV_{sphere} \Rightarrow$$

$$N^1\big((X', S', t + \Delta t) \leftarrow (X, S, t)\big) = \frac{g(S, S')(1 - e^{-\mu_s c \Delta t})e^{-\mu_a c \Delta t}}{\frac{3\sqrt{2}c\Delta t(1 - S * S')}{4dr} + 1} N(X, S, t) \quad (17)$$

## 3. Photon density calculation and System propagation matrix construction

To formulate the Dynamic Radiative Transfer System (DRTS) we have to consider the numerical calculation procedure and subsequent application in computing. Thus, we need to partition the time, the travelling direction and the Euclidean space. For calculation purposes time can be partitioned in small equally spaced steps $\Delta t$, fulfilling the condition $\mu_s c \Delta t \ll 1$, which is in agreement with the above theoretical analysis.

In the case of biological tissue, because of high scattering coefficient, the time partitioning will be in the femtosecond regime. This naturally makes the DRTS method applicable in modeling photon propagation from the femtosecond up to steady state. We emphasize mostly in the femtosecond regime, where this method has obvious advantages, which will be discussed in a later section.

In principle, any reasonable numerical recipe for partitioning the spatial (Euclidean space) can be chosen, for example tetrahedral, cubic volumes etc. The same fact holds for the angular partitioning, for example partitioning based on icosahedron, Lebedev points etc. In general, one can use any simplex and appropriate base function over the space $R^3 \times S^2$.

In order to simplify and explicate the implementation of DRTS method, a close packing scheme of equal spheres was selected for spatial partition, which is close to the previous theoretical analysis. For the same reasons, Lebedev points [37] were chosen as directions for the angular partition. Let's define $X_i$ to be the centers of the spheres and $S_n$ be the directions. Since partition spheres are of equal volumes, the number of photons in each sphere $(X_i, dr)$ at time $t$ directed to $S_n$ represents the photon density $J(X_i, S_n, t)$, which for simplicity has a uniform distribution inside each sphere. Notice that conversion to light intensity (energy) can be done simply by multiplying with $E = hf$.

In the case of population $N^0$ the photons at time $t + \Delta t$ will reach a sphere with center $B$ according to the previous theoretical analysis. But a center $B$ will in most cases lie between other centers $X_j$ which were defined according to the partition of the Euclidean space. A first order approximation can be established by the distance of the nearest-neighbor centers $X_j$ to center $B$. So the equation of photon density for the non-scattered population will be:

$$J^0\left((X_j, S_n, t + \Delta t) \leftarrow (X_i, S_n, t)\right) = w^0_{jn,in} J(X_i, S_n, t)$$
$$\text{where:} \quad w^0_{jn,in} = \frac{(2dr - \|X_j - B\|)}{\sum_k^{\|X_k - B\| < 2dr}(2dr - \|X_k - B\|)} e^{-\mu_a c \Delta t} e^{-\mu_s c \Delta t} \quad (18)$$

for every $X_k$ such that $\|X_k - B\| < 2dr$ and $X_j \in \{X_k\}$, where $B = X_i + c\Delta t\, S_n$

A similar approach is used for the scattered population $N^1$. We apply a first order approximation using nearest-neighbor centers $X_j$ around line segment $AB$, weighted proportionally to their distance from $AB$. So the equation of photon density for the scattered population will be:

$$J^1\left((X_j, S_m, t + \Delta t) \leftarrow (X_i, S_n, t)\right) = w^1_{jm,in} J(X_i, S_n, t)$$

where:

$$w^1_{jm,in} = \frac{2dr - d(X_j, AB)}{\sum_k^{d(X_k, AB) < 2dr}(2dr - d(X_k, AB))} g(S_n, S_m)(1 - e^{-\mu_s c \Delta t}) e^{-\mu_a c \Delta t} \quad (19)$$

$AB$ is line segment, $A = X_i + c\Delta t\, S_m$, $B = X_i + c\Delta t\, S_n$
$d(X_k, AB)$ is the distance of a center $X_k$ from line segment $AB$
for every $X_k$ such that $d(X_k, AB) < 2dr$ and $X_j \in \{X_k\}$

Equations (18), (19) express the photon contribution from a sphere-direction $(X_i, S_n)$ at time $t$ to another sphere-direction $(X_j, S_m)$ at $t + \Delta t$. The part of photon contribution into $(X_j, S_m, t + \Delta t)$ originating from $(X_i, S_n, t)$ will be:

$$J\left((X_j, S_m, t + \Delta t) \leftarrow (X_i, S_n, t)\right) = a_{jm,in} J(X_i, S_n, t) \quad (20)$$

$$a_{jm,in} = \begin{cases} w^1_{jm,in}, & \text{when } S_n \neq S_m \\ (w^0_{jn,in} + w^1_{jm,in}), & \text{when } S_n = S_m \\ 0 & \text{when nearest neighborhood} \\ & \text{conditions do not hold} \end{cases} \quad (21)$$

The addition in the second term of the previous equation (when $S_n = S_m$) includes the contribution of forward scattering effects at zero angle.

To organize calculations we construct a state vector $\boldsymbol{J}(t)$ with elements $J(X_i, S_n, t)$. This vector conveys all possible information about the state of the system at the corresponding time instant. The dimension of $\boldsymbol{J}$ is:

$$D(\boldsymbol{J}) = (\text{number of spatial partition points}) \cdot (\text{number of angular partition directions}) \quad (22)$$

We now construct a *square system evolution matrix* $\boldsymbol{A}$, of dimension $D(\boldsymbol{J}) \times D(\boldsymbol{J})$, which, when multiplied by the state vector $\boldsymbol{J}(t)$, will give the new state vector $\boldsymbol{J}(t + \Delta t)$:

$$\boldsymbol{J}(t + \Delta t) = \boldsymbol{A}\,\boldsymbol{J}(t) \quad (23)$$

The elements of $\boldsymbol{A}$ are $a_{jm,in}$ given from equation (21). This matrix conveys all possible information about the time evolution of photon propagation inside a particular media.

$\boldsymbol{A}$ is a probability matrix expressing the medium optical behavior, time-independent, which needs to be computed only once, independent of source properties or other external parameters.

By factoring out the absorption coefficient from each element of the matrix $\boldsymbol{A}$ the above equation can be written as:

$$\boldsymbol{J}(t + \Delta t) = e^{-\mu_a c \Delta t} \boldsymbol{A_S}\, \boldsymbol{J}(t) \quad (24)$$

where $\boldsymbol{A_S}$ has information only about the scattering phenomenon, thus we can call it system scattering matrix.

Since the factor $e^{-\mu_a c \Delta t}$ is constant, matrix $\boldsymbol{A_S}$ characterizes the system. The fact that light can travel only a finite distance $c\Delta t$ (given that the media being partitioned has bigger dimensions than $c\Delta t$), means that only a few line and column elements are non-zero, as there is "no connection" (no possibility of any photon to propagate) between remote spatial points in that time interval.

We can summarize some important properties for both matrices $\boldsymbol{A}$ and $\boldsymbol{A_S}$:

- Matrices $\boldsymbol{A}$ and $\boldsymbol{A_S}$ are sparse. This is easily seen since every point-direction element can be affected only by others that lie within the sphere $(\boldsymbol{X}, c\Delta t)$.

- The summation of every column of $\boldsymbol{A_S}$ is 1 (because of population conservation).

- Every number in $\boldsymbol{A}$ and $\boldsymbol{A_S}$ is positive less or equal to one, or zero ($0 \leq A_{ij} \leq 1$). This comes from the fact every element in arrays express a probability.

- The diagonal is zero. Diagonal elements correspond to photons returning at the same center and same direction. But after time $\Delta t$ no photons will be in the same place except photons that have backward direction that travel $c\Delta t/2$ and return, albeit in the opposite direction. This can also be verified by the equation (13) that relates $\boldsymbol{S}$ and $\boldsymbol{S}'$.

- The matrices are irreducible under the condition the volume is connected and every center $X_i$ has at least one direction $S_j$ such that the line segment $X_iX_k, where\ X_k = X_i + \frac{3}{2}c\Delta tS_j,$ lies inside the volume of interest.

To prove the last one let's create a directed graph with vertices $J_{X_iS_j}$ where the directed edges from $J_{X_iS_j}$ to $J_{X_kS_l}$ exist if there is a contribution of photons from $(X_i, S_j)$ to $(X_k, S_l)$; this means that the corresponding value in the matrix is positive. Matrices $A$ and $A_S$ are irreducible if and only if there is a path from every vertex to any vertex, or the graph is strongly connected. Starting from a vertex $J_{X_iS_j}$ under the first condition there should be a point $X_m = X_i + c\Delta tS_j$ inside the volume. According to the distribution of photon population of type 1 there will be some photons that are scattered at center $X_i$ from any direction $S_0$ to direction $S_j$. This population will reach point $X_m$ at time $t + \Delta t$.. From vertex $J_{X_mS_j}$ photons travelling to the direction $S_j$, which are scattered to $-S_j$ at point $X_k = X_m + \frac{1}{2}c\Delta tS_j,$ will reach the same point $X_m$ but with opposite direction. Now from point $X_m$ with direction $-S_j$ photons will reach the starting point $X_k$ and scattered to any direction $S_l$. Thus, the vertices $J_{X_iS_j}$ and $J_{X_iS_l}$ are connected.

So starting from vertex $J_{X_iS_0}$ we can reach every vertex of type $J_{X_iS_l}$. The transition path is

$$J_{X_iS_0} \xrightarrow{g(S_j,S_0)} J_{X_m,S_j} \xrightarrow{g(-S_j,S_j)} J_{X_m,-S_j} \xrightarrow{g(S_l,-S_j)} J_{X_iS_l} \qquad (25)$$

Since by starting from a center and specific direction we can always reach any direction of the same center, now it suffices to prove that starting from a center $X_i$ we can reach any center $X_l$. This can be proved by the following procedure.

From center $X_i$ there is a path to every center that lies in sphere($X_i$, $c\Delta t$). If the center $X_l$ has been reached the proof is completed, otherwise we "mark" the boundary centers, remove the remaining centers inside the sphere and repeat the procedure for the marked centers. Since the volume is connected and the remaining volume becomes smaller at every step, the algorithm will terminate and will reach center $X_l$. Therefore this concludes the sought proof that matrices $A$ and $A_S$ are irreducible.

## 4. An approach to modeling the effects of the sources and boundary conditions in media with multiple refractive indices

A general dynamic system calculation method like DRTS in order to be meaningful, it should be universally applicable in a multitude of physical phenomena. In our case this includes boundary conditions, different media, sources and phenomena like specular or diffusive reflection, refractive deflection, etc. and, finally integrating the measurement process for these phenomena. In this section we outline the approach for introducing all relevant phenomena and processes in the DRTS methodology in a consistent way. The analytic expression in each case depends on the mathematical description or modeling and on the detailed properties of the examined materials, measurement probes, etc.

### 4.1. The effect of the sources

Typically sources of light are described with a time and space – dependent power flow (intensity) function having a defined volume or surface distribution. To incorporate this information in our dynamic system modeling (DRTS) it must first be expressed as a spatial – directional distribution of photons for every single time step $\Delta t$. This is necessary as our measure $J$ counts the number of photons inside a volume of a sphere $(X, dr)$ at distinct points of time $k\Delta t$, where conversion of energy to number of photons can be done using $E = h\nu$. Using the previously defined space and direction partitions, it is possible to translate the source influence into photon-population contributions inside the spheres and angular directions. To calculate the contribution of sources, the number of photons that are produced or inserted between in the time interval $[k\Delta t, (k+1)\Delta t]$ has to be expressed. Also the position and direction of them at time $(k+1)\Delta t$ have to be determined. These source contributions, because of the discretized system description, will be expressed by a correlation matrix of photon presence inside the specific spheres and direction angles due to their outflow from the sources. Therefore in the general case our system will be described by a system:

$$J(t + \Delta t) = A\, J(t) + B\, P(X, S, t) \tag{26}$$

Where $P(X, S, t)$ is the discrete source vector of photon production, and $B$ is the correlation matrix with our positional centers $X_i$, directions $S_j$ and time. Here, for an accurate calculation, the time-partition of the source function should be implemented in shorter time intervals $dr/c$.

### 4.2. Media boundaries and measurement processes

In this section the processes related to photons crossing boundaries between different physical media as well as the final measurement of the emerging photon populations will be described. In the most complicated scenario the media will differ in their absorption and diffusion parameters and refractive index. The treatment now involves a 2 – step process for the photons passing through the boundary within the time interval $\Delta t$. During the first step the propagation is calculated until the photon reaches the media interface in time shorter than $\Delta t$. At this point the relevant photon population is considered as a new Dirac source for the second medium and its propagation for the remaining time (till the end of interval $\Delta t$) is similarly evaluated using the new medium parameters. This method has the advantage that it does not essentially modify the previous methodology but simply makes elements of $A$ matrix more complicated to evaluate. It should be noted, however, that this evaluation should be performed only once.

Photons passing through a surface with a change of velocity will produce specular reflection and deflection. These phenomena must be included in our calculations.

Since the problem is transformed to geometrical problem only a brief outline of the solution will be given. Photons will travel over the path $X_0 C$ in accordance to the equations given in previous section by replacing the path $c\Delta t$ with the actual one $\|X_0 C\|$. At point $C$ two new sources will be produced, one at direction $S^r$ that will obey the reflection lows and one at $S^t$ that will obey refraction lows. The effect produced at time $t + \Delta t$ from these sources can be easily calculated since they are of Dirac type. In the second figure is it shown that the part $\frac{\|C'C\|}{\|X_0 C\|}$ of population $N_1$ that was turned from $S_0$ to $S$ will be distributed to line segment $X'X''$ with direction $S'$ because of reflection and the transmitted part will be distributed over $X^o X''$ with direction $S^o$. The absorption of this transmitted part will be proportional to the path the photons travel inside the volume. This can be easily calculated, because of linearity, by computing the effects at the extreme points $X^o$ and $X''$. Population type $N_1$ produced by the Dirac type source after reflection may hit one more time the surface depending on the direction of a possible subsequent scattering. Since at very short time intervals only one change of direction is possible, there will be no further hits on the media interface in time $\Delta t$. So these two steps conclude the photon propagation calculation giving again geometrical interpretations to the dynamic system evolution.

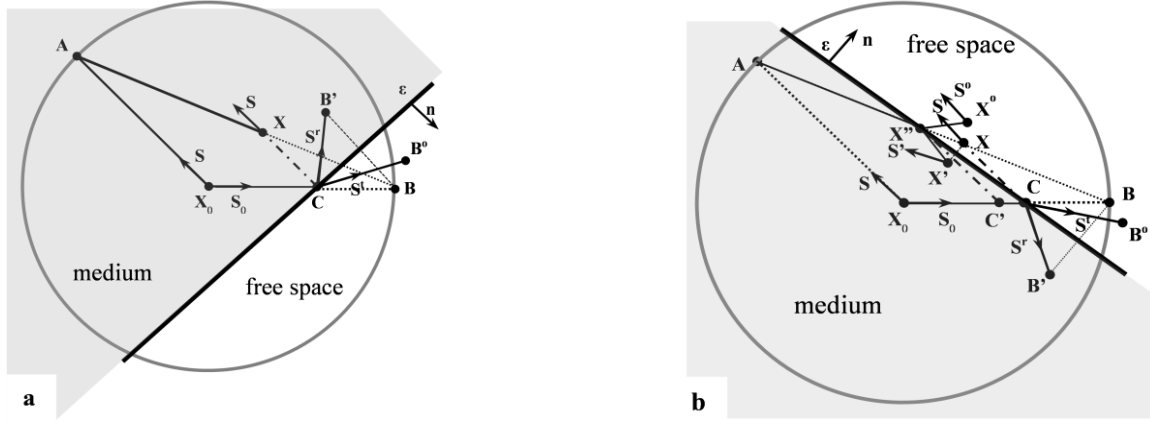

**Figure 2**. Photon population position after time $\Delta t$ when reaching surface of different media
**a**. At point $C$ two rays will be created, reflected $S^r$ and transmitted $S^t$.
**b**. Reflection of population $N_1$, population will be distributed at $X'X''$ (reflected) and $X''X^o$ (transmitted).

Measurements especially in a laboratory environment involve finite volumes so boundary effects have to be investigated in accordance with a possible measurement system. Sensors of random photons are based on photon-count (as PMT, Photodiode) or in energy absorption (thermal-type), however both types can be treated as photon-count sensors. On the other hand any sensor response is based on a time integral even if the time interval is very small. The response of any measurement system, which is in a particular position in space inside the volume of interest will be proportional to photons reaching the sensor in a certain time–window. A system will respond with a measure $M(t)$ to the number of photons $P^M$ that arrive at the sensor according to a temporal and directional response sensitivity $G(X,S)h(t)$, where $h(t)$ is the impulse response of the sensor.

$$M(t) = \iint_{s,t-w}^{t+w} G(X,S)h(t-\tau)P^M(\tau)\,d\tau ds \cong \sum_{t-w}^{t+w} F(X_i^o, S_j, t-k\Delta t) J(X_i^M, S_j, k\Delta t) \qquad (27)$$

From previous analysis the photon-distribution information is described by $J(t)$. Specifically $J(X_i^M, S_j, k\Delta t)$ is the part of $J(t)$ at the sensor location and $F(X_i^o, S_j, k\Delta t)$ represents the impulse response of positional and directional sensitivity, according to our spatial and angular partition.

In the real-life case that the sensor resides outside the volume of interest, one can simply extend the partitioned volume. This will not introduce any significant computational cost, because there is no scattering or absorption events in free space and the photons simply travel along their direction.

## 5. Simulation Results

The algorithm from the previous theoretical analysis is summarized according to the following flowchart (*Figure 3*).

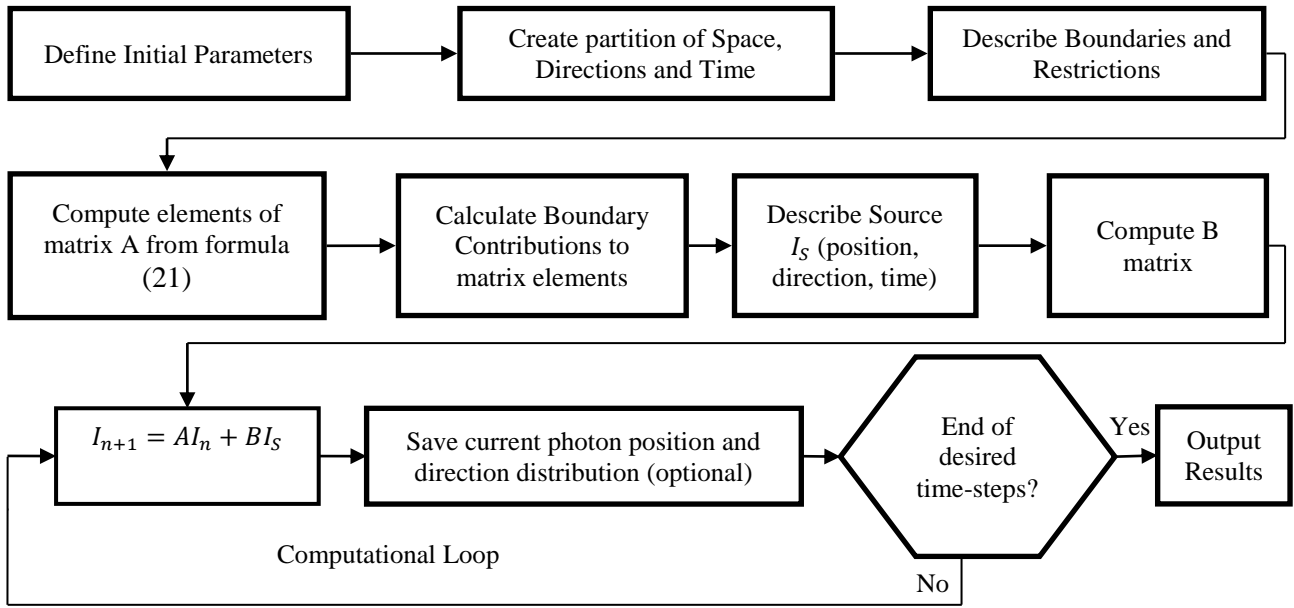

*Figure 3. Flowchart of the DRTS algorithm*

This algorithm was implemented in C# and was applied for the calculation of propagation of ultra-fast photon pulses in various scattering and absorbing media. For testing purposes, initial simulations assumed an optically uniform medium shaped as rectangular box. The light source is a collimated directional beam propagating along the y-axis ("horizontal" axis in our figures and video) impinging at the center of the y=0 plane. The partition of the media volume was $100 \times 100 \times 100$ close-packing of equal spheres. Directional partition was implemented with the Lebedev rule using 86 directions. To fulfill the requirements of equation (5) and according to the optical properties of the medium ($\mu_a, \mu_s$), a suitable time step had to be defined. In our simulations $\Delta t = 0.1/(c \cdot max\{\mu_a, \mu_s\})$ was selected. The angular scattering dependence of single scattering events, was approximated by the Henyey-Greenstein scattering function. In the simulation results below, two cases are presented, one for isotropic scattering ($g = 0$) and one for the forward scattering ($g = 0.7$), with the latter being a typical value for biological tissues. The spatial dimension of the input photon source here occupies a "cuboctahedron" volume of 13 neighboring spheres (a central sphere plus 12 closest-neighbor spheres) positioned at the center of the $y = 0$ face of the medium. A number of photons are emitted at $t = 0$ from all source cells along the y-direction, simulating a single-directional delta-function distribution over time. This source should be viewed as an ultra-short incoherent light pulse, in order to avoid complications arising from interference phenomena, which have not yet been included in our model. Also, for simplicity, no reflection effects from the medium external surfaces were considered in theses simulation runs. Numerical results are depicted below in figures and graphs, where all data are represented in logarithmic scales because of the extreme spread in data values, typically $10^{20}$.

## 5.1. Simulating strongly scattering, weakly absorbing, isotropic homogeneous media

For a strongly scattering and weakly absorbing medium ($\mu_s/\mu_a = 100$) with isotropic scattering, the total photon density distribution for three different time frames ($t_1 = 2\Delta t, t_2 = 6\Delta t, t_3 = 15\Delta t$) is shown in *Figure 4* sequence. It should be noted that all calculations were done in 3 spatial dimensions, from which these 2D graphs were subsequently extracted for the sake of clarity of representation. Photon densities are represented with pseudo-colors covering a $10^{20}$ dynamic range according to the explanatory logarithmic color scale at the right side.

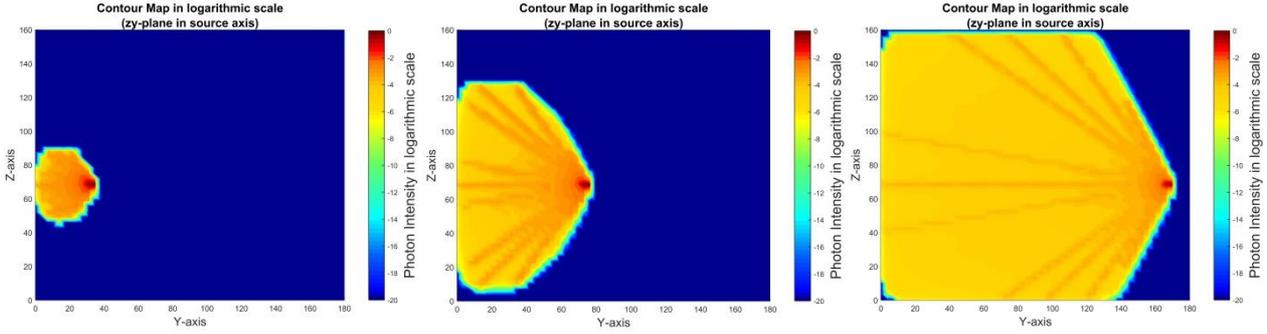

*Figure 4. Photon Density in logarithmic scale for a diffusing light pulse inside strongly scattering media and anisotropy factor $g = 0$, at three different time frames ($t_1 = 2\Delta t, t_2 = 6\Delta t, t_3 = 15\Delta t$). Notice that the radial-type lines are simulation artifacts, due to angular partition.*

The red tip of the distribution corresponds to un-scattered photons traveling at the speed of light ($c=c_o/n$) in the medium, while other parts of the distribution correspond to scattered photons towards all possible directions. The line-patterns inside the distribution (in the form of diverging "fan-out" trails) are simulation artifacts due to the finite number of directions (86) considered in our angular-space partition.

A full 3D frame of the same sequence is included here in **Figure 5**, to demonstrate the shape of the propagating photon distribution as it diffuses in the scattering medium.

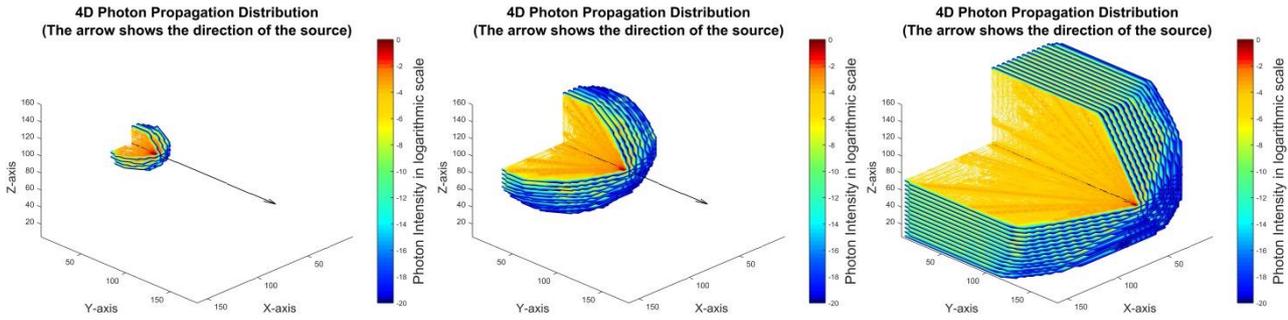

*Figure 5. Time propagation of Photon Density in 3D for a diffusing light pulse inside strongly scattering media, The arrow indicates the source direction, in three different time frames ($t_1 = 2\Delta t, t_2 = 6\Delta t, t_3 = 15\Delta t$).*

The following **Figure 6** and **Figure 7** sequences show the spatial distribution of photons traveling along a single direction, specifically either along the original source y-direction (0º) or perpendicular to that (+90º angle), in the same material as above.

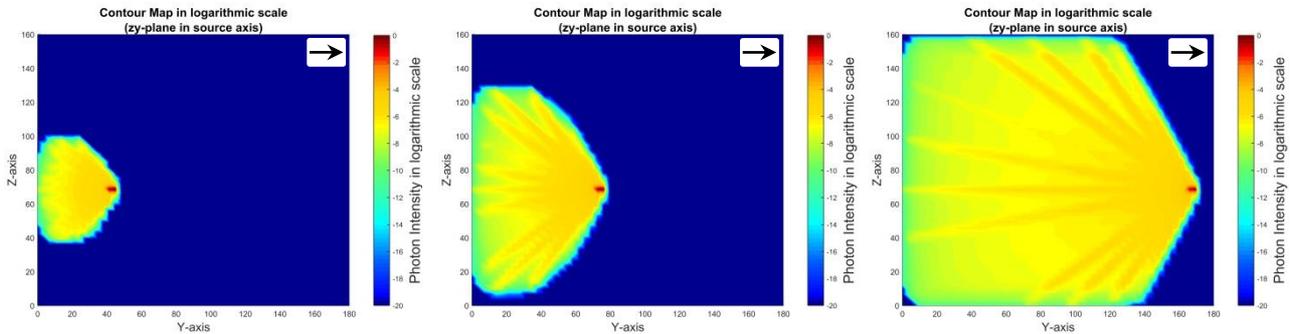

*Figure 6. Photon populations traveling exclusively along the original source direction (y-axis) - indicated by the arrow at top right corner - at three different time frames ($t_1 = 2\Delta t, t_2 = 6\Delta t, t_3 = 15\Delta t$).*

In **Figure 6** sequence we have isolated the photon populations traveling exclusively along the y-axis at each time frame (indicated by arrow at top right corner of each contour map image). These photons happen to be either yet unscattered (population at the front tip of the distribution), or have gone through a suitable number of scatterings (at least 2) that allowed them to resume in the y-axis direction.

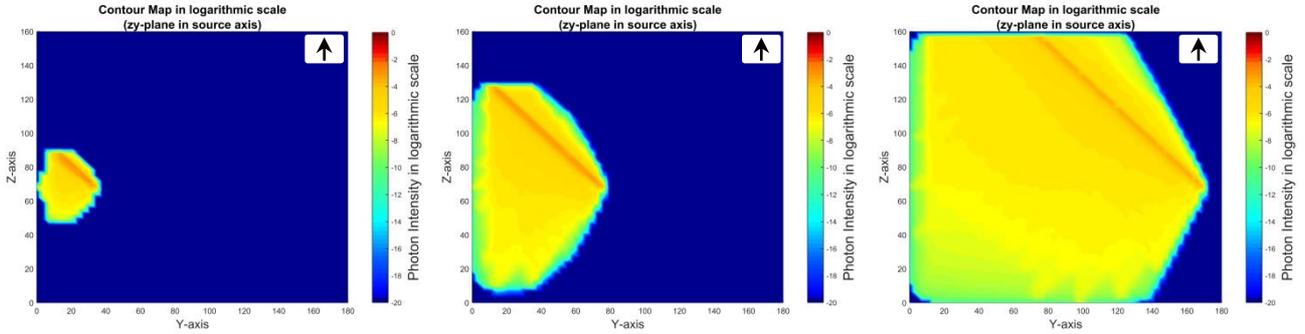

*Figure 7. Photon populations traveling along the positive z-axis (i.e. at 90º to the original source y-direction) - indicated by the arrow at top right corner - at three different time frames ($t_1 = 2\Delta t, t_2 = 6\Delta t, t_3 = 15\Delta t$).*

Correspondingly, in **Figure 7** sequence we have isolated, at each time frame, the photon populations traveling along the +z-axis, perpendicular to the original y-axis source direction (indicated by the arrow in top right corner of each contour map image). These photons have undergone a suitable number of scatterings (at least one) that streamed them in the +90º direction. The peak of the photon density is a line representing photons that had undergone only one scattering. This line distribution was expected from the theoretical analysis of section 2, and illustrated in **Figure 1**.

### 5.2. Simulating strongly forward scattering, weakly absorbing, isotropic homogeneous media

For a medium with a different forward scattering (g=0.7) and same $\mu_a, \mu_s$ coefficients as before, the photon propagation is plotted in **Figure 8** sequence. The contour map shows, as expected, increased density values in directions closer to the direction of the source.

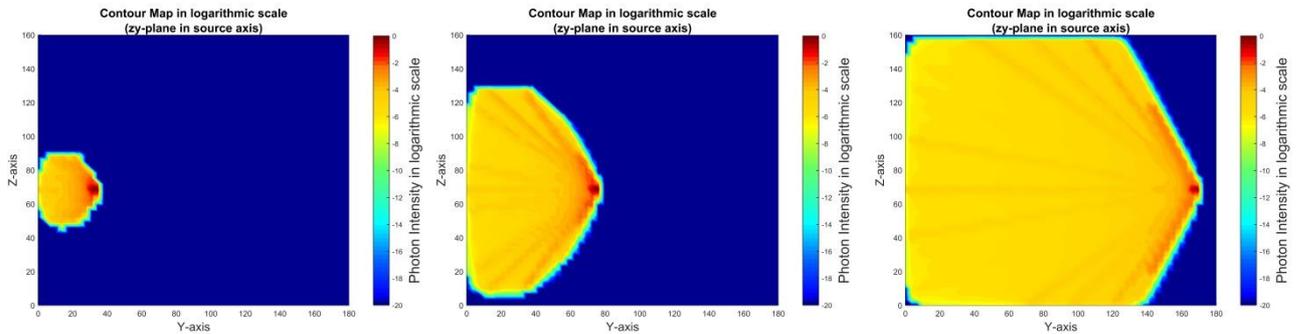

*Figure 8. Photon populations for a diffusing light pulse inside strongly scattering media with anisotropy factor g=0.7, at three different time frames ($t_1 = 2\Delta t, t_2 = 6\Delta t, t_3 = 15\Delta t$). The density in directions closer to the direction of the source is increased compared to Figure 4(red area around wave front).*

### 5.3. Simulating strongly scattering, strongly absorbing, isotropic homogeneous media

For a higher absorption coefficient value ($\mu_\alpha/\mu_s = 4$) and isotropic scattering medium, the analogous propagation of total photon population distribution at different times is shown in the picture sequence of **Figure 9**.

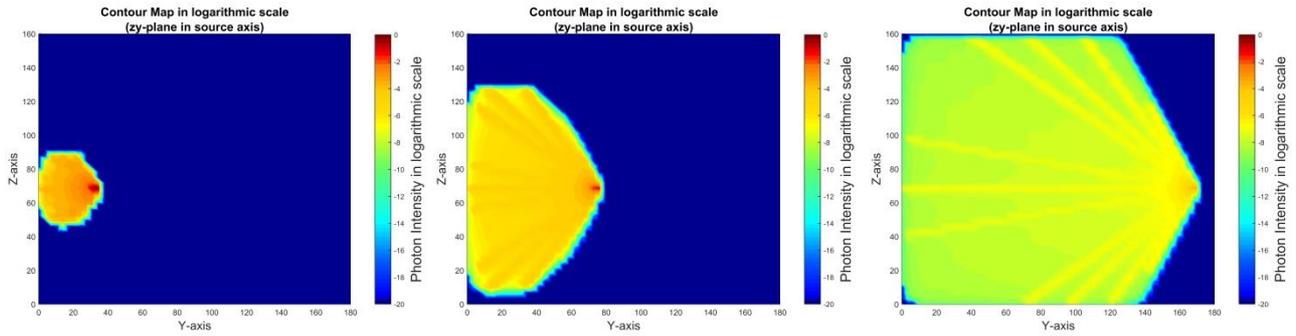

*Figure 9.* Photon populations for a diffusing light pulse inside strongly absorbing media with isotropic scattering g=0, at three different time frames ($t_1 = 2\Delta t, t_2 = 6\Delta t, t_3 = 15\Delta t$).

### 5.4. Simulating a measurement system

As explained in section 4.2, in equation (**27**), a measurement system will respond to photon population, time interval and the characteristics of the sensor. Since in our simulation no special sensor was assumed, a hypothetical response of a single-directional point detector is presented in **Figure 10**.

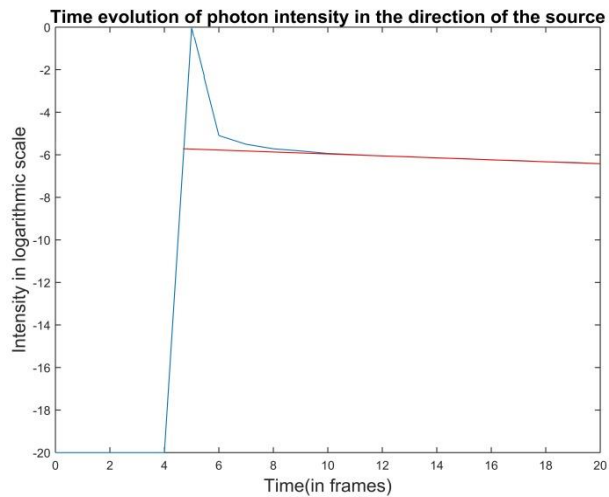

*Figure 10.* Time evolution of photon intensity at a point lying along the source direction. The red line indicates the intergo-diferrential RTE solution, which cannot describe the spike produced by the light front at the sensor position. The blue line indicates the DRTS solution, which accurately represents the actual physical phenomenon.

The hypothetical sensor was positioned on the source direction axis at a distance of 1/5 of the media length and assumed to respond only to photons traveling along that direction. The graph shows the surge of the light-front intensity arriving at the sensor on the fifth time step, corresponding to unscattered photons. Subsequently, the diffused photon population arrives, with the time-distribution gradually tailing-off exponentially. This signal decay is indicated by the straight line (red line in figure) of the log plot, the slope of which determines the combined absorption and scattering coefficients, therefore it can be further used to derive the light propagation characteristics of the medium. It should be noted that intergo-diferrential RTE cannot describe the spike produced by the light-front arrival.

In comparison with previous simulation results on similar problems [38], our Fig. 11 graph results are in agreement, though it must be noted that those refer to quite specific approximations and conditions (e.g. forward scattering, bio-tissue optical media, 1-dimensional), while our DRTS results are not limited in any way by such constraints. Regarding our theoretical model analysis (Sections 2 – 4) and ultrafast 3-D propagation simulations (Sections 5.1 – 5.4) above, these appear for the first time in scientific literature at the time of writing, to the best of our knowledge.

## 6. Discussion: Comparisons and Conclusions

The new DRTS method was developed as a mathematical and computational tool suitable for solving important physical problems of time-evolution of photon propagation in absorptive and scattering media. Thus, it is interesting to discuss this method in comparison with other approximation methods used to solve RTE, such as FEM-type methods and stochastic methods such as Monte Carlo. Both DRTS and RTE try to express mathematically the same physical phenomena but they differ in their philosophy. Specifically, RTE uses the conventional interpretation of physical phenomenon and vector space by approaching the concept of variation of quantity and space-point via limit to zero and arrives to an integro – differential equation. Then, for calculation purposes, the solution space of the integro – differential expression is approximated by an infinite functional base. But in a real measurement system neither zero nor infinity can be reached, so large errors are quite possible. DRTS tries to avoid these two dangerous domains (limit to zero and infinity) by proceeding via a dynamical system approach, where each system state-vector leads via the global system-evolution matrix with linear operations to the next system state, in a deterministic manner. On the other hand, both Monte Carlo and DRTS methods employ the same, exact statistical description of the physical phenomenon, where however DRTS is based on a probability-matrix approach resembling a Markov process. Therefore MC and DRTS greatly differ in calculative resource requirements and, as consequence, in their capabilities and adaptability to solving demanding problems. Another important difference between MC and DRTS approaches lies in their "global view" of the solution. Specifically, DRTS stores the global material-volume information in a unique matrix-form that is calculated only once. In contrast, MC is not "aware" of the universal system behavior and attempts to resolve the full solution by "throwing dices" and obtaining numerous individual-case solutions, then summing the partial pieces of information, so that the expected value will be approximated after a large number of iterations. Thus, the MC methodology makes it particularly tedious and resource-demanding to study fast time-dependent phenomena. In contrast, DRTS is based on a pre-determined and time-independent system matrix, thus DRTS can easily handle such phenomena.

Another interesting comparison between RTE and DRTS can be made on the basis of the flexibility aspects. For example, approximation methods that attempt to solve RTE generally cannot handle in a universal fashion reflection, refraction, non-convex volume shapes (involving re-entry of radiation), etc. By contrast, as previously indicated in section 4, DRTS can handle these phenomena and solve these problems under a general and consistent scheme, since DRTS intrinsically incorporates these effects in the original single calculation of the system matrix.

Comparing the resource requirements and computational cost, advantages of DRTS become evident, once the general flow scheme *Figure 3* is considered. The method starts with the calculation of the elements of system matrix A (equation (21)), a time- and computation-demanding step, which however will have to be performed (and stored) only once for the whole process. It should also be pointed out that this system matrix **A** is sparse, a fact that greatly reduces problem complexity, storage requirements and computational cost for the subsequent steps. To calculate the photon-system status on the next step after time $\Delta t$ has elapsed, it suffices to simply multiply the above system matrix **A** with the initial system status-vector and add any new source term (equation (26)). The new vector result, after $\Delta t$, can be reused on-the-fly in RAM as the starting point for the subsequent step, where matrix A will remain un-altered. All subsequent computational cost will only involve elementary sparse-matrix – vector calculations (without any matrix inversions), all with non-negative numbers. Thus there is absolutely no cause for calculation instabilities.

To demonstrate the superiority of the DRTS method versus the mainstream standard calculation methods (Monte Carlo and Finite Elements – like methods) we consider their comparative practical application in simulations of similar complexity as the test-case problems of the previous section. In the these examples, DRTS was programmed and run on a standard PC machine and handled easily the ultra-short light-pulse propagation where photon density was described by a $86 \cdot 10^6$– elements vector, producing a variety of useful results for any desired time sequence without any instabilities or storage overflow problems.

An equivalent computation by means of Monte-Carlo method, for comparison, would be greatly inferior. For the same number of partition micro-volumes and for the given dynamic range, the required number of iterations with Monte-Carlo algorithms would be at least a hundred times the dynamic range of photon density. In our method the precision of calculation is restricted only by computer precision. In our simulation results a dynamic range of $10^{20}$ is presented. For this range Monte-Carlo would need at least $10^{22}$ iterations, which is prohibitively large on standard computer systems. Also worth noting, on every iteration MC has to produce random numbers and calculate exponential or logarithmic probabilities. Such dynamic range is crucial when modeling transient phenomena, as is exactly the case with ultrafast light pulse diffusive propagation. Additionally, as Monte-Carlo methods require summation of all the individual intermediate results to give the final answer, all these data would have to be stored, handled and retrieved, requiring an increase of required system memory. For our particular example, memory requirements for MC would be $86 \cdot 10^6 * time\ steps * 10 bytes \approx [time\ steps] * Gb$ (since double precision uses 10bytes of memory). It should be noted that any intermediate results in DRTS are used just once and storage to disk is performed only if results output is desired at that time step, as also indicated in the flowchart. Overall, even for relatively simple problems of low complexity, as presented above, Monte-Carlo calculations demand vast computational costs, related to the required iteration numbers and execution time, and also to the extreme storage requirements.

In comparison with integro-differential models, an equivalent simulation with finite elements would involve a similar number of iterations as with the DRTS above, albeit including matrix-inversion operations at each step, therefore vastly increasing the computational cost. This is due to the fact that approximation methods assume that the solution to a differential-equation system can be described by a weighted summation of a base functions set, the coefficients of which have to be determined. This leads to a linear equation system, which requires a matrix inversion in every time step. Apart from the computational cost, a harmful side-effect is the potential non-positivity of the solution, which is physically non-acceptable in the case of radiation intensity.

Another computational advantage of DRTS over both MC techniques and RTE approximations relates to the re-evaluation of system response in cases of modifying the source properties or configuration. Previous methods have to fully recalculate everything from scratch if any modification is made in the original source positions, directions or time-dependences. On the other hand, in DRTS array ***A*** remains unchanged, and – as typically a light beam occupies very confined spatial dimensions – array ***B*** will be non-zero only for the corresponding elements. The only modification therefore will be array ***B*** and the source vector, equation (**26**), so the computational cost in DRTS again is comparatively much smaller.

According to the previous analysis, DRTS presents a novel perspective of thought, a new theory and a concrete method for treating radiation-propagation problems, especially for the ultrafast-source and transient phenomena time regimes, which offers several important advantages:

- The new method is able to accurately predict the time-dependent evolution of photon flow, even for fast-varying sources. This is important in solving the Optical Tomography "forward problem" in the case of femtosecond pulsed excitation.

- The freedom of extension the volume of interest (as in the case of media boundaries and measurement) gives the ability to overcome mathematical difficulties caused by non-convex volumes. In general DRTS is not affected if the volume is enclosed inside a convex one (like a simple cube or a sphere) and simply solves the problem over the new extended regime.

- DRTS uses a dynamical-system formality and provides physical and geometrical interpretations for the system evolution. The basic structure of the method can be enhanced to accept more complicated problems which are closer to the physical reality.

- The DRTS method allows the calculation either of photon population-density or the derivation of directional flow. This can be achieved by simple matrix operations.

Further work will be directed towards DRTS error analysis, detailed modeling of pulsed sources, interference phenomena and application-software development.